\documentclass[aps,prl,showpacs,twocolumn]{revtex4}
\usepackage{graphics,wrapfig,times}
\usepackage{graphicx}
\usepackage{amsmath}
\usepackage{amssymb}

\newcommand{\bvec}[1]{{\mathbf #1}}

\newcommand{\be}{\begin{equation}}
\newcommand{\ee}{\end{equation}}
\newcommand{\bea}{\begin{eqnarray}}
\newcommand{\eea}{\end{eqnarray}}

\usepackage{color}

\begin{document}

\title{Birefringent break up of Dirac fermions in a square optical lattice}
\author{Malcolm P. Kennett, Nazanin Komeilizadeh, Kamran Kaveh, and Peter M. Smith}  
\affiliation{Physics Department, Simon Fraser University, 8888 University Drive, Burnaby, British Columbia, V5A 1S6, Canada}
\date{\today}

\begin{abstract}
 We generalize a proposal by S{\o}rensen {\it et al.} [Phys. Rev.
Lett. {\bf 94}, 086803 (2005)] for creating an artificial magnetic
field in a cold atom system on a square optical lattice. 
This leads us to an effective lattice model
with tunable spatially periodic modulation of the artificial magnetic field and the hopping
amplitude.  When there is an average flux of half a flux quantum per plaquette
the spectrum of low-energy excitations 
 can be described by massless Dirac fermions in which the usually doubly degenerate 
Dirac cones split into cones with different ``speeds of light'' which can be tuned to give a single
Dirac cone and a flat band.  These gapless birefringent Dirac
fermions arise because of broken chiral symmetry in the kinetic energy term of the effective low
energy Hamiltonian. 
We characterize the effects of various
perturbations to the low-energy spectrum, including staggered potentials, interactions, and
domain wall topological defects.
\end{abstract}


\maketitle

With the discovery of graphene \cite{graphene} and topological insulators \cite{TI}
there has been much recent interest in systems in which low energy
excitations can be described using Dirac fermions.
A parallel area of interest has been the exploration of the possibility of generating
artificial magnetic fields for cold atoms confined in an optical lattice.  
Neutral bosonic cold atoms cannot couple to
a magnetic field directly, so there have been numerous
proposals \cite{Cooper,Sorensen,Proposals,Lim}
of approaches to couple atoms to an artificial magnetic field, several of which
have been implemented experimentally \cite{Cornell,Spielman}.

The problem of the spectrum of quantum particles in a 
uniform magnetic field on a lattice has the well-known Hofstadter
spectrum \cite{Hofstadter}.
Our modification of the proposal by S{\o}rensen {\it et al.} \cite{Sorensen}
leads to an effective Hamiltonian with a
 tunable Hofstadter-like spectrum that arises from the 
combination of hopping and an artificial magnetic field 
with a non-zero average that are both periodically modulated in the
$x$ {\it and} the $y$ directions.  The presence of spatial 
periodicity in the  \emph{amplitude} as well as the phase
of the hopping is the key difference between the model we
consider here and previous work on the spectrum of particles 
in the presence of magnetic fields that  are periodic in both 
the $x$ and $y$ directions \cite{Barelli}.  This difference 
facilitates the unusual Dirac-like spectrum that we discuss 
in this Letter.

In our effective model, when there is an average of
half a flux quantum per plaquette, and at half-filling, 
the low energy degrees of freedom can be described 
by a Dirac Hamiltonian with the unusual property that chiral symmetry 
is broken in the kinetic energy rather than via mass terms.    
This has the consequence that the doubly degenerate Dirac cone for massless 
fermions splits into two cones with tunable distinct slopes, analagous to a situation
 in which there are two speeds of light for fermionic excitations,
similar to birefringence of light in crystals such as calcite.  We discuss the meaning of 
broken chiral symmetry in our effective model and  
explore the effects of various perturbations,
such as staggered potentials, domain walls, and interactions between
fermions.  

The approach to obtain an artifical magnetic field for cold 
atoms in an optical lattice suggested by S{\o}rensen {\it et al.} \cite{Sorensen}
was presented in the context of the Bose-Hubbard model, but ignores
interactions and is not specific to bosons.  We
consider a model of spinless fermions (corresponding to only
one available hyperfine state for cold atoms) with Hamiltonian

\begin{eqnarray}
H=-{\mathcal J}
\sum_{\langle i,j\rangle}(\hat{c}^\dagger_i \hat{c}_j + \hat{c}^\dagger_j
 \hat{c}_i) ,
\label{eq:BHM}
\end{eqnarray}
where $\hat{c}_i^\dagger$ and $\hat{c}_i$ are fermionic creation and annihilation
operators respectively at site $i$, $\hat{n}_i= \hat{c}_i^\dagger \hat{c}_i$ 
is the number operator, and the notation $\langle i,j \rangle$ indicates that we
restrict the sum in the hopping term to nearest neighbours only.  
There can be no Hubbard-like interaction for spinless fermions, and since 
nearest neighbour interactions in an optical lattice system are weak, we
postpone our discussion of interactions.

In Ref.~\cite{Sorensen}, two steps are required to generate an 
artificial magnetic field.  First, a time-varying quadrupolar potential 
$V(t) = V_{qp} \sin(\omega t) \hat{x} \hat{y}$ is applied to the system, 
and second, the hopping is modulated as a function of time.  During the 
course of one oscillation of the quadrupolar potential, hopping 
in the $x$ direction is turned for a very
short period of time $\tau \ll t_0 = \frac{2\pi}{\omega}$ 
at times $t = nt_0$, where $n$ is an integer, and 
hopping in the $y$ direction is turned on for time $\tau$ around 
$t = \left(n+\frac{1}{2}\right)t_0$.  Due to
the periodic oscillation in the Hamiltonian, 
the time evolution operator after $m$ periods may be written as
$U(t=mt_0) = U(t=t_0)^m$. 

\begin{figure}[htb]
\includegraphics[width=6cm]{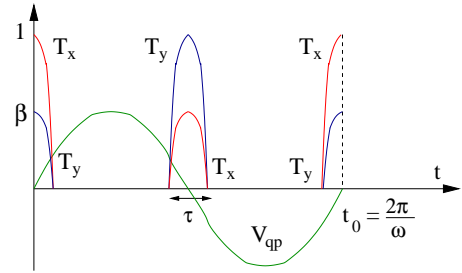}
\caption{Time dependence of the hopping and the quadrupolar potential
during the course of one period of the quadrupolar potential.}
\label{fig:hopping}
\end{figure}

Our modification to the proposal in Ref.~\cite{Sorensen} 
is that when hopping is turned on in the $x$-direction at time $t = nt_0$, hopping is also turned on in 
the $y$-direction with an amplitude $ 0 \leq \beta \leq 1$ relative to the hopping in the $x$-direction.  
At time $t = \left(n+\frac{1}{2}\right)t_0$, hopping is turned on in the $y$ direction, 
and hopping in the $x$-direction is 
turned on with amplitude $\beta$ relative to the hopping in the $x$-direction as
illustrated in Fig.~\ref{fig:hopping}.

\begin{figure}[htb]
\hspace*{0.9cm}\includegraphics[width=8.5cm]{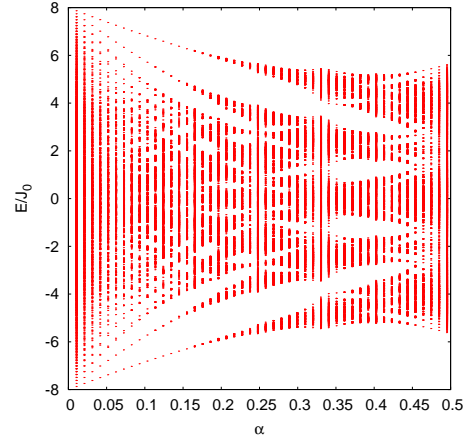}
\caption{Spectrum as a function of $\alpha$ when $\beta = 1$ obtained by exact
diagonalization on a 97$\times$97 site lattice: there is no artificial magnetic field, 
yet due to the periodic hopping, the spectrum has some similarities with 
the Hofstadter spectrum.}
\label{fig:beta1}
\end{figure}

The operators for hopping in the $x$ and $y$ directions, are 
$\hat{T}_x  =  -{\mathcal J} \sum_{x,y} \left(\hat{c}^\dagger_{x+1,y} \hat{c}_{x,y} + {\rm h.c.} \right)$ 
and 
$\hat{T}_y  =  -{\mathcal J} \sum_{x,y} \left(\hat{c}^\dagger_{x,y+1} \hat{c}_{x,y} + {\rm h.c.} \right)$ 
respectively,
and we may write the time evolution operator as
\begin{eqnarray} 
U\left(t = mt_0\right) &  = & \left[ e^{-\frac{i\tau}{2\hbar}\left(\hat{T}_x + \beta 
\hat{T}_y\right)} e^{2\pi i \alpha \hat{x}\hat{y}}
e^{-\frac{i\tau}{\hbar}\left(\beta \hat{T}_x + \hat{T}_y\right)} \right. \nonumber \\
& & \left. \hspace*{1cm} \times  e^{-2\pi i \alpha \hat{x}\hat{y}} 
e^{-\frac{i\tau}{2\hbar}  \left(\hat{T}_x + \beta \hat{T}_y\right)}\right]^m ,
\end{eqnarray}
where $\alpha = V_{qp}/\pi\hbar\omega$ and we have set the lattice constant to unity. 
To lowest order in ${\mathcal J}\tau/\hbar$ we can write this in the form
$$ U = e^{-\frac{i H_{\rm eff} t}{\hbar}},$$
where the effective Hamiltonian is
\begin{eqnarray}
H_{\rm eff}&  = & - J_0 \sum_{x,y} \left\{ \left[ \left(1 + \beta e^{2\pi i \alpha x}\right) \hat{c}^\dagger_{x,y+1}\hat{c}_{x,y} + \,  {\rm h.c.} \right]
\right. \nonumber \\
& & \left. \hspace*{0.6cm}
 + \left[\left(\beta + e^{2\pi i \alpha y}\right) \hat{c}^\dagger_{x+1,y} \hat{c}_{x,y} + \, {\rm h.c.}\right]\right\},
\label{eq:Heff}
\end{eqnarray}
with $J_0 =  \tau {\mathcal J}/t_0$.  A more conventional way to write this Hamiltonian is in the form
\begin{eqnarray}
H_{\rm eff} = -\sum_{ij} \left[t_{ij} e^{\frac{ie}{\hbar} \int_j^i \bvec{A}\cdot d\bvec{l}} \hat{c}^\dagger_i \hat{c}_j + \, {\rm h.c.}\right] ,
\end{eqnarray}
from which we may identify the amplitude of the hopping:
\begin{eqnarray}
t_{x+1,y} &  =  & J_0 \sqrt{1 + \beta^2 + 2\beta \cos(2\pi \alpha y)}, \\
t_{x,y+1} &  =  & J_0 \sqrt{1 + \beta^2 + 2\beta \cos(2\pi \alpha x)} ,
\end{eqnarray}
and the artificial magnetic field
\begin{eqnarray}
B_z & = & \frac{2\pi\alpha \hbar}{e} \left\{ \frac{\beta^2 + \beta \cos(2\pi\alpha x)}{1 + \beta^2 + 2\beta \cos(2\pi\alpha x)}
\right. \nonumber \\ & & \hspace*{1cm} \left.
 - \frac{1 + \beta \cos(2\pi \alpha y)}{1 + \beta^2 + 2\beta \cos(2\pi \alpha y)}\right\}.
\end{eqnarray}
This field is the sum of a spatially uniform piece with magnitude 
$\frac{2\pi\hbar \alpha}{e}$ and a piece that is  spatially periodic 
in both the $x$ and $y$ directions. If $\beta = 0$,  the hopping amplitude
is $J_0$ and the field is uniform with strength 
$\frac{2\pi\hbar \alpha}{e}$, corresponding to a
flux of $\alpha \phi_0$ per plaquette (where $\phi_0$ is the flux
quantum) as found in Ref.~\cite{Sorensen} and there is a Hofstadter spectrum.
If $\beta = 1$, then $B_z = 0$, but the hopping parameters are still 
spatially periodic.
At $\beta$ intermediate between 0 and 1, 
both the hopping and the magnetic field are spatially periodic in $x$ and $y$.  
This illustrates the essential difference between the model we consider and
previous work on quantum particles in a periodic magnetic field on a lattice 
-- there is spatial periodicity of $1/\alpha$ in the amplitude of the hopping as well as
 in the magnetic field.  For finite $\beta$ the spectrum (illustrated for $\beta = 1$
in Fig.~\ref{fig:beta1}) as 
 a function of $\alpha$ is  reminiscent of the Hofstadter spectrum.

\begin{figure}[htb]
\includegraphics[width=6cm]{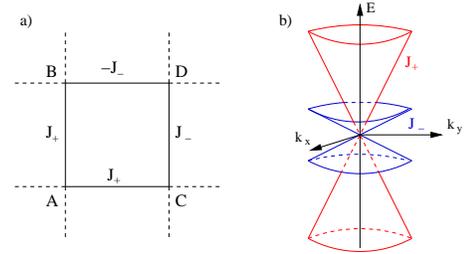}
\caption{a) Unit cell of tight binding model with hopping parameters indicated.
b) Dirac cones corresponding to $J_+$ and $J_-$ bands.}
\label{fig:hop}
\end{figure}

When $\alpha = 1/2$ there is an average of half a flux quantum per 
plaquette, and the
theory is time reversal symmetric 
as fermions cannot detect the sign of the flux \cite{Seradjeh}.
The effective Hamiltonian Eq.~(\ref{eq:Heff}) simplifies to a tight
binding model with four
sites in the unit cell as shown in Fig.~\ref{fig:hop} a). 

Labelling the four sites in the unit cell as $A$, $B$, $C$, and $D$, and Fourier transforming in space, 
we may rewrite the effective Hamiltonian in the 
following form:
\begin{eqnarray}
\label{model}
H = \sum_{\bvec{k}} \psi_k^*[E_k - {\mathcal H}_k] \psi_k ,
\end{eqnarray}
with
$${\mathcal H}_k = 2\left(\begin{array}{cccc} 0 & J_+\cos k_y & J_+ \cos k_x & 0 \\
 J_+ \cos k_y & 0 & 0 & -J_- \cos k_x \\ J_+ \cos k_x & 0 & 0 & J_- \cos k_y \\
0 & -J_- \cos k_x & J_- \cos k_y & 0 \end{array} \right),$$  
where $J_\pm = J_0(1 \pm \beta)$ and (in row vector form)
$ \psi_k = (c_{Ak}, c_{Bk}, c_{Ck}, c_{Dk}).$
We find that the dispersion is 
$$ E_{k} = \pm J_\pm \sqrt{\cos^2 k_x + \cos^2 k_y} , $$
and we note that when $\beta = 1$, $J_- = 0$, so there will be a 
flat band at $\epsilon = 0$ and a dispersing band associated with $J_+$.
We can also see that in the vicinity of the points
$\bvec{K}_{\pm,\pm} = \left(\pm\frac{\pi}{2},\pm\frac{\pi}{2}\right)$,
the spectrum is linear
\begin{eqnarray}
E_{q} = \pm J_\pm \sqrt{ q_x^2 + q_y^2} ,
\label{eq:twocones}
\end{eqnarray}
where $\bvec{q} = \bvec{k} - \left(\pm\frac{\pi}{2},\pm\frac{\pi}{2}\right)$,
 and there are cones with two
different slopes, corresponding to $J_\pm$ respectively as illustrated in 
Fig.~\ref{fig:hop} b).  
When $\beta = 0$, the two slopes are identical, whereas as
$\beta \to 1$, the $J_-$ band becomes flat, and the $J_+$ band
remains as a cone.  
Several authors recently considered lattice models for cold atoms 
that are equivalent to the $\beta = 1$ limit of our model, in which 
there are three bands, one flat, and one Dirac like \cite{Shen,Apaja}.  When
$\beta \neq 1$, the underlying Dirac structure of the problem 
is exposed, allowing us to understand this unusual dispersion from a 
symmetry point of view \cite{footnote}. 

To start in this direction, we expand around the Dirac points and
 represent the low energy theory  (with $\bvec{k}$ measured with respect
to $\bvec{K}$)
\begin{eqnarray} 
H_k = 2J_0\left[\left(\gamma^0 \gamma^1 + i\beta\gamma^3\right)k_x 
+ \left(\gamma^0 \gamma^2 + i\beta \gamma^5\right)k_y\right],
\end{eqnarray}
where we use a non-standard representation of the gamma matrices in 
which $\gamma^0 = \sigma_3 \otimes \sigma_3$, $\gamma^1 = i\sigma_2 
\otimes I_2$, $\gamma^2 = i\sigma_3 \otimes \sigma_2$, 
$\gamma^3 = -i\sigma_1\otimes I_2$, and $\gamma^5 = -\gamma^0 \gamma^1 \gamma^2 \gamma^3
 = -i\sigma_3\otimes \sigma_1$.
The matrices $\gamma^0$, $\gamma^1$, $\gamma^2$ and $\gamma^3$
satisfy the Clifford algebra $\gamma^\mu \gamma^\nu + \gamma^\nu 
\gamma^\mu = 2g^{\mu\nu}$ with Minkowski metric $g^{\mu\nu}$. 

The dimension of the minimal representation of the Clifford algebra in 
2+1 dimensions is 2, allowing for the
$2\times 2$ Pauli matrices as a choice for the $\gamma$s. A non-minimal 
$4\times 4$ representation as we have used above leads to a freedom in the choice of the 
$\gamma^{0}$ matrix, i.e. a matrix with $\left(\gamma^0\right)^2 = I_4$ that anticommutes
with $\gamma^{1}$ and $\gamma^{2}$.  Candidates for $\gamma^{0}$ are then $\{ \gamma^{0}, 
\gamma^{0}\gamma^{3},\gamma^{0}\gamma^{5}, \gamma^{1}\gamma^{2}\}.$
The matrices $\{\gamma^{0}, \gamma^{0}\gamma^{3}, \gamma^0 \gamma^{5} \}$ form a 
triplet and $\gamma^1 \gamma^2$ forms a singlet with respect to the SU(2) 
``chiral''-symmetry group with generators $\{ \frac{i}{2}\gamma^{3},
\frac{i}{2}\gamma^{5}, \frac{i}{2}\gamma^{35}\}$ (where $\gamma^{35} \equiv  \gamma^{3}\gamma^{5}$).
Each different choice of $\gamma_{0}$ corresponds to a different labelling of the four sites in the 
unit cell.  The elements of the chiral group  generate transformations between each labelling. 
For example, the generator $\gamma^{5}$ translates the plaquette indices to the 
labelling of the neighboring lattice cell along the $y$-direction, 
whilst $\gamma^{3}$ translates the plaquette indices to the neighbouring cell in the $x$-direction.

\be
e^{\frac{\pi}{2} \gamma^{5}} \left(\begin{array}{cc}c_{A}\\c_{B}\\ c_{C}\\ c_{D}\end{array}\right) = 
 i\left(\begin{array}{cc} c_{B}\\ c_{A}\\ -c_{D}\\-c_{C} \end{array}\right),
e^{\frac{\pi}{2}\gamma^{3}} \left(\begin{array}{cc}c_{A}\\c_{B}\\ c_{C}\\ c_{D}\end{array}\right) = 
 i\left(\begin{array}{cc} c_{C}\\ c_{D}\\ c_{A}\\ c_{B} \end{array}\right). 
\ee
Similarly, $\gamma^{35}$ translates the plaquette one lattice cell along the $x$- and one lattice
cell along the $y$- direction.
When $\beta = 0$, the elements of the chiral group are symmetries of $H_k$.

When $\beta \neq 0$, the $\gamma^3$ and $\gamma^5$ terms in $H_k$ break the
chiral symmetry and shifts along either the $x$- or $y$-directions do not leave $H_k$ invariant.
 This manifest chiral symmetry breaking is inherently different from the conventional notion of
spontaneous chiral symmetry
breaking in field theoretical models which is the signature of mass generation \cite{Miransky}.

An additional discrete symmetry of the $H_k$ (that arises from 
the hopping structure in ${\mathcal H}_k$) that holds even 
when $\beta \neq 0$ is
$$ \Gamma = \frac{i}{2}\left(\gamma^1 \gamma^3 + \gamma^2 \gamma^5
\right) - \frac{i}{2}\left(\gamma^2\gamma^3 - \gamma^1 \gamma^5
\right),$$
which corresponds to a reflection about the diagonal $AD$ in the 
unit cell, with $c_A \to c_A$, $c_B \to c_C$, $c_C \to c_B$ and 
$c_D \to -c_D$.  The action of $\Gamma$ on $H_k$ is to exchange $k_x$ and
$k_y$. 

{\it Fermion birefringence}: As illustrated in Fig.~\ref{fig:hop} b)
the dispersion Eq.~(\ref{eq:twocones}) 
admits massless fermions with two different ``speeds of light''
controlled by $\beta$.  
The eigenvectors (written as row vectors) 
for the positive and negative energy $J_+$ bands
are 
$ \Psi_1 = \frac{1}{\sqrt{2}}\left(1,  -\sin\theta,
 - \cos\theta, 0\right)$ and
$\Psi_2 = \frac{1}{\sqrt{2}}\left(1,\sin\theta, \cos\theta, 0\right)$;
whilst the eigenvectors for the $J_-$ bands are
$\Psi_3 = \frac{1}{\sqrt{2}}\left(0, \cos\theta,  - \sin\theta, 1\right)$ and
$\Psi_4 = \frac{1}{\sqrt{2}}\left(0, -\cos\theta, \sin\theta, 1\right),$
where we write $k_x = k\cos\theta$ and $k_y = k\sin\theta$.
The linear combinations $\Psi_1 + \Psi_2$ and $\Psi_3 + \Psi_4$
have non-zero amplitude only on $A$ and $D$ sites respectively.
Any other state
 will break up into fast ($J_+$) and slow ($J_-$) fermionic 
excitations, analogous to fast and slow modes in a birefringent medium. 

{\it Staggered potentials}:  Staggered on-site potentials
are a natural perturbation to $H_k$
in the context of cold atoms on an optical lattice.  We can write the most general form of 
such a potential as 
\begin{eqnarray}
\Delta & = & \sum_k \psi_k^\dagger \left[\Delta_0 I_4 + \Delta_1 \gamma^0 + \Delta_2 (i\gamma^1  \gamma^3 
+ i\gamma^2 \gamma^5) \right. \nonumber \\ 
& & \left. \hspace*{1cm} + \Delta_3 (i\gamma^1 \gamma^3 - i\gamma^2 \gamma^5)\right] \psi_k,
\end{eqnarray}
where we may set $\Delta_0 = 0$ since this 
just corresponds to a uniform shift of the chemical potential.
The $\Delta_{1}$ term violates chiral symmetry in the usual way but is Lorentz 
invariant and hence introduces a gap in the dispersion of the fermions 
\begin{eqnarray}
E_k = \pm \sqrt{\Delta_1^2 + 4 J_\pm^2 k^2}.
\label{eq:delta1}
\end{eqnarray}
When $\beta = 1$ there are flat bands 
at $E = \pm \Delta_1$ that intersect the $J_+$ bands only at $(k_x, k_y) = (0,0)$.
The birefringence property discussed above is unaffected by the $\Delta_1$ term.
We combine $i\gamma^{1}\gamma^{3}$ and $i\gamma^{2}\gamma^{5}$ into a Lorentz
invariant term ($\Delta_2$) and a Lorentz violating term ($\Delta_3$). 
There are two cases in which we have
obtained simple analytic solutions for the spectrum:
case I): $\Delta_1 \neq 0$, $\Delta_2 \neq 0$, $\Delta_3 = 0$, for which
$$ E_k = \left\{\begin{array}{c} \Delta_2 \pm \sqrt{(\Delta_1 + 
\Delta_2)^2 + 4J_+^2 k^2} \\ -\Delta_2 \pm \sqrt{(\Delta_1 - 
\Delta_2)^2 + 4J_-^2 k^2} \end{array} \right. ,$$
and case II):
 $\Delta_1 \neq 0$, $\Delta_2 = 0$, $\Delta_3 \neq 0$, for which
$$ E_k = \left\{\begin{array}{c} \Delta_3 \pm \sqrt{(\Delta_1 -
\Delta_3)^2 + 4J_+^2 k_y^2 + 4J_-^2 k_x^2} \\
 - \Delta_3 \pm \sqrt{(\Delta_1 + \Delta_3)^2 + 4J_+^2 k_x^2
 + 4J_-^2 k_y^2} \end{array} \right.  .$$
In case I) the dispersion remains isotropic in momentum space and there
are flat bands when $\beta = 1$, whereas in
case II), the dispersion becomes anisotropic, with the anisotropy governed
by $\beta$ through $J_\pm$.
In both cases, there is a shift in the spectrum
and there will be at least one set of massive modes (however in both cases 
there can be a set of massless modes whose dispersion is given by 
the upper half of a cone if  
$\Delta_1 = \pm \Delta_{2,3}$ and $\Delta_0 = \mp \Delta_{2,3}$).  

{\it Interactions}: as we consider spinless fermions, there
will be no on-site Hubbard interaction, so we 
consider  nearest neighbour interactions of the extended
Hubbard type (for cold atoms in an optical lattice 
these will generally be weak):
\begin{equation}
H_{\rm int} = \sum_{\left<ij\right>} V_{ij} n_i n_j.
\end{equation}
Setting all of the $V_{ij} = V_0$,
we can write the interaction Hamiltonian in terms of
spinors as

\begin{eqnarray}
H_{\rm int} = \frac{V_0}{16}\sum_k \left[(\bar{\psi}_k\gamma^0\psi_k)^{2} - 
(\bar{\psi}_k\psi_k)^{2} \right] ,
\end{eqnarray}
with $\bar{\psi}_k = \psi_k^\dagger \gamma^0$.  The identity and
$\gamma^0$ that appear in the kernels of the quartic
interaction terms are the only elements of the Clifford algebra 
that either commute or anticommute with all of the elements of the
Lorentz group and the chiral group, ensuring that the interactions
remain invariant under any rotation of the lattice by the Lorentz
group or relabelling of the plaquette indices by the chiral group.
At the mean field level, the $(\bar{\psi}\psi)^{2}$ term breaks 
the chiral symmetry by introducing an effective mass term 
$m_0\gamma^0$, and the $(\bar{\psi}\gamma^{0}\psi)^{2}$ term
renormalizes the chemical potential as $\delta I_4$ and
is otherwise uninteresting.  
In the limit of weak interactions, the mean field interaction
Hamiltonian is:

\begin{eqnarray}
H_{\rm int}^{\rm MF} & = &  \sum_k \psi^\dagger_k \left[(\delta I_4 + m_0 \gamma^0)
 + (m_1 \gamma^0 \gamma^1 + m_2 \gamma^0 \gamma^2 \right. \nonumber \\
& & \left. \hspace*{1cm} + m_3 i\gamma^3 + m_5 i\gamma^5) \right]\psi_k  ,
\end{eqnarray}
where $\delta = \left< n_{A}\right> + \left< n_{B}\right> + \left< n_{C}\right> + \left< n_{D}\right>$, 
and the order parameter for staggered charge density wave order 
$m_0 = \left< n_{A}\right> - \left< n_{B}\right> - \left< n_{C}\right> + \left< n_{D}\right>$
arise from the Hartree term. The remaining masses, $m_1$, $m_2$, $m_3$ and $m_5$ arise from the Fock term 
-- if these are dropped and $\beta = 0$, we recover the mean-field approximation of the Gross-Neveu 
model \cite{GrossNeveu}.
Similarly to a $\Delta_1 \gamma^0$
staggered potential, the Hartree term leads to massive excitations,
but does not destroy fermion birefringence.  The detailed study of interactions when
$\beta \neq 0$ is a topic for future investigation.

For small values of $\beta$, when $H_{\rm int}$ is added to Eq.~(\ref{eq:Heff})
there is a mapping between the weak interaction strength regime considered above to the
strong interaction strength limit that preserves the property of birefringence:
\be
E_{k}(\beta, V_{0}) = \beta E_{k}(\beta^{-1}, \beta^{-1}V_{0}) .
\ee
 This arises from the appearance of the chiral symmetry
generators, $\gamma^{3}$ and $\gamma^{5}$ in the kinetic energy 
and their duality with Lorenz group generators
$\gamma^{0}\gamma^{1}, \gamma^{0}\gamma^{2}$. 
Upon choosing a different representation of Clifford algebra elements, 
one can transform
$\gamma^{0}\gamma^{1} \leftrightarrow \gamma^{3}, \gamma^{0}\gamma^{2}\leftrightarrow \gamma^{5}$. 

{\it Topological defects}: broken chiral symmetry at $\beta \neq 0$ implies that there cannot be 
vortices, but
domain walls of the form $\Delta_1(x) \gamma^0$ where
$\lim_{x \to \infty} \Delta_1(x) = \Delta$ and $\lim_{x \to -\infty}
\Delta_1(x) = -\Delta$ can occur.   If $\beta = 0$ 
then the form of the solutions with energy $|\epsilon| < \Delta$ is
well known.  When $\beta \neq 0$ we can find zero energy bound states
with different spatial extents for the + and - solutions:
$$ \psi_+(x) = e^{-\kappa_+ \int_0^x ds \Delta_1(s)} u_+; \quad 
\psi_-(x) = e^{-\kappa_- \int_0^x ds \Delta_1(s)} u_-,$$
where $u_+ = (1, 0, i, 0)$, $u_- = (0,-i,0,1)$, and
$\kappa_\pm = 1/2J_\pm$. 

In this Letter we have demonstrated a model 
whose low energy excitations are birefringent fermions 
that arise from broken chiral symmetry. We discuss
the low energy properties of the model and illustrate
the meaning of broken chiral symmetry in our model.
We argue that such a model could be realised by cold atoms 
in an optical lattice.  This might not be the only route  
-- as noted in a similar context in Ref.~\cite{Seradjeh} 
another approach might be through an appropriately engineered 
semiconductor heterostructure.

An important feature of the birefringent fermion dispersion that we find here
is that the slopes of the $J_+$ and $J_-$ bands can be controlled by the parameter 
$\beta$.  In particular when $\beta = 1$, there can be flat bands in the spectrum 
and these flat bands are robust to the addition of a staggered potential 
$\Delta_1 \gamma^0$ and weak nearest neighbour Hubbard 
interactions at the Hartree level.  Flat bands 
such as Landau levels can lead to interesting correlated phases when interactions
beyond mean field are taken into account \cite{Green}.  This suggests that 
future avenues for research on this model could include the study of 
such correlated phases when $\beta \neq 0$, and the generalization of the model to
fermions with spin.  Including spin would allow for on-site Hubbard interactions, which 
would be considerably complicate matters and require techniques similar to 
those that have been used to study high temperature superconductors \cite{Herbut}.

The authors acknowledge very helpful discussions with Igor Herbut and Bitan Roy. 
We acknowledge support from NSERC and the 
use of Westgrid computing facilities.


\begin{thebibliography}{99}

\bibitem{graphene} K.S. Novoselov, {\it et al.}, Science {\bf 306}, 666 (2004);
K.S. Novoselov, {\it et al.}, Nature {\bf 438}, 197 (2005); G. W. Semenoff,
 Phys. Rev. Lett. {\bf 53}, 2449 (1984).

\bibitem{TI} C. L. Kane and E. J. Mele, Phys. Rev. Lett. {\bf 95}, 146802 (2005);
D. Hsieh, {\it et al.}, Nature {\bf 452}, 970 (2008).

\bibitem{Cooper} N. R. Cooper, Adv. Phys. {\bf 57}, 539 (2008).

\bibitem{Proposals} N. R. Cooper and N. K. Wilkin, Phys. Rev. B {\bf 60},
 16279 (1999); 
D. Jaksch and P. Zoller, New. J. Phys. {\bf 5}, 56 (2003); 
 A. Klein and D. Jaksch, Europhys. Lett. {\bf 85}, 13001 
(2009); M. Rosenkranz, A. Klein, and D. Jaksch, arXiv:0909.0728v1;
F. Gerbier and J. Dalibard, New. J. Phys. {\bf 12}, 033007 (2010);
T.-L. Ho and S. Zhang, arXiv:1007.0650v1; D. Makogon, I. B. Spielman, and
C. Morais Smith, arXiv:1007.0782v1.

\bibitem{Lim} L.-K. Lim, A. Hemmerich, and C. Morais Smith, Phys. Rev. A
{\bf 81}, 023404 (2010).

\bibitem{Sorensen}  A. S. S{\o}rensen, 
E. Demler, and M. D. Lukin, Phys. Rev. Lett. {\bf 94}, 086803 (2005).

\bibitem{Cornell}  V. Schweikhard, {\it et al.}, 
Phys. Rev. Lett. {\bf 92}, 040404 (2004).

\bibitem{Spielman} Y.-J. Lin, {\it et al.}, Nature {\bf 462}, 628 (2009).

\bibitem{Hofstadter} D. R. Hofstadter, Phys. Rev. B {\bf 14}, 2239 (1976).

\bibitem{Barelli} A. Barelli, J. Bellissard, and R. Rammal, J. Phys. France
{\bf 51}, 2167 (1990);
M. C. Chang and Q. Niu, Phys. Rev. B {\bf 50}, 10843 (1994);
G.-Y. Oh, Phys. Rev. B {\bf 60}, 1939 (1999); 
M. Ando, {\it et al.}, J. Phys. Soc. Jpn. {\bf 68}, 3462 (1999);
M.-C. Chang and M.-F. Yang, Phys. Rev. B {\bf 69}, 115108 (2004);
Y. Iye, {\it et al.}   Phys. Rev. B {\bf 70}, 144524 (2004);
Y.-F. Wang and C.-D. Gong, Phys. Rev. B {\bf 74}, 193301  (2006).

\bibitem{Seradjeh} B. Seradjeh, C. Weeks, and M. Franz, Phys. Rev. B {\bf 77}, 033104 (2008).

\bibitem{Shen} R. Shen, {\it et al.}, Phys. Rev. B {\bf 81}, 041410(R) (2010).

\bibitem{Apaja} V. Apaja, M. Hyk\"{a}s, amd M. Manninen, Phys. Rev. A
{\bf 82}, 041402(R) (2010).

\bibitem{footnote} Ref.~\cite{Seradjeh} demonstrates Dirac points in a 
square lattice model closely related to our model -- the structure of the
hopping is such that it is not left-right and up-down symmetric when 
$\bvec{m} \neq 0$ which leads to a gap in the spectrum, in contrast to 
our model in which hopping is left-right and up-down symmetric.

\bibitem{Miransky} V.A.Miransky, {\it Dynamical Symmetry Breaking in 
Quantum Field Theories} (World Scientific, Singapore, 1993).

\bibitem{GrossNeveu} D. J. Gross and A. Neveu, Phys. Rev. D {\bf 10}, 3235 (1974).

\bibitem{Green} D. Green, L. Santos, and C. Chamon, Phys. Rev. B {\bf 82}, 075104 (2010)

\bibitem{Herbut} I. F. Herbut, Phys. Rev. B {\bf 66}, 094504 (2002).

\end{thebibliography}
\end{document}